# Superresolution technique beyond the diffraction limit under a structured beam via different optical nanostructures

Subhankar Roy[1], Jianping Hu[2] and M Ummal Momeen[1*]

*[1]Magnetic Instrumentation and Applied Optics Laboratory, Department of Physics, School of Advanced Sciences, Vellore Institute of Technology, Vellore, Tamil Nadu, 632014, India.*

*[2]Applied Nuclear Technology in Geosciences Key Laboratory of Sichuan Province, Chengdu University of Technology, Chengdu, People's Republic of China.*

**ummalmomeen@gmail.com*

***ABSTRACT:***

To overcome the limit of diffraction while achieving the superresolution technique, solid immersion lenses are the key optical elements for data storage and nanophotonics applications. Recent demonstrations have shown how different nanostructures (such as elliptical solid immersion lenses) are used in diverse fields of increasing resolution in the presence of a structured Gaussian beam. By applying twisted beams such as angular momentum beams (Laguerre- Gaussian) and spatial higher-order Gaussian beams (Hermite- Gauss), we can attain a sharp near-field focal spot pattern, which is considerably better than the conventional solid immersion lens structure in mm scale specifically for imaging beyond diffraction limit. Our computation results present a resolution of 27 nm under a specific Hermite -Gauss mode illumination on a pyramidal shape nanolens structure. By numerical simulations, tolerance has been confirmed with a slight variation in beam size and geometrical modification to make the model compatible with fabrication errors. This narrow bandwidth intensity distribution can be utilized for scanning the sample with higher resolution, especially in the field of quantum technology.

***Introduction:***

In optical microscopy, fluorescence-based light manipulation to acquire resolution in the subwavelength regime is quite challenging. Surpassing the classical far field Abbey diffraction limit to attain superresolution in the subwavelength region is highly demanding for resolving nanoscale particles/specimens in optics. The classical diffraction limit significantly restricts the resolution between adjacent particles while scanning through a sample because of the exponential decay of evanescent waves, which carry subwavelength information in the far-field domain [**1**]. Near-field focal spot with high lateral resolution can be used to analyze subwavelength structures such as tapered fibers, optical nanoantennas, and nanoscale needles through the coupling of light near the surface of a sample [**2, 3**].

Initially, the design of the super lens was potentially capable of retrieving the evanescent wave to achieve higher resolution in imaging [**4**]. The design of such a lens effectively converts evanescent waves into propagating waves in the near-field region, as they are engineered by a negative index material. Few optical superlenses are fabricated from silver or SiC, which works well in the infrared and ultraviolet wavelength regions [**5,6**]. On the other hand, silver metal substantially supports evanescent wave transport without decaying much inside the silver slab, which effectively carries information [**4**]. However, these optical systems suffer from few drawbacks, which substantially hinder further applications. These lenses generate heat while scanning the sample, which in turn damage and affects the target material during optical imaging under white light illumination [**7**]. Hyperlenses were proposed and fabricated to increase the resolution by using metamaterials, which also suffer from inevitable manufacturing limitations [**8**]. To overcome this, nanorange lenses (solid immersion lenses (SILs)) with dielectric materials have been developed, which undergo low loss and surpass the classical diffraction limit barrier for subwavelength imaging. These nanorange models are able to produce better resolution than the normal macroscopic SIL [**9**]. Even the inclusion of high-refractive-index materials can immensely increase the resolution below 100 nm [**9-10**]. Solid immersion microscope uses near field mode and it utilizes the evanescent field at the interface of lens material, which can be applied for near field imaging. We slightly modified the hemispherical shape into ellipsoidal shape, which can be called as ellipsoid SIL structure (e-SIL). This structure devises new set of degrees of freedom as an optimization parameter, which even can provide a better tuning/control over simulation results alongside a tight near field focal spot [**9, 11].**

Nevertheless, out of several optical approaches [**12, 13]**, SIL has been extensively studied because of its exceptional performance in improving the superresolution ranging from the micrometer size to the nanoscale regime [**14**]. These SILs show efficient development in terms of near-field confinement under different kinds of wave illuminations, such as plane wave, Gaussian wave, and dark field incidence, beyond the diffraction limit **[9-15]**. Different vectorial beams such as Laguerre Gauss, Bessel Gauss is adopted to attain a resolution beyond diffraction limit **[16,17]** where azimuthal polarized beam can generate much smaller focal spot with incoherence superposition analogy **[18]**. These spatially resolved structured beam is able to produce strong interference at the focus from different part of the beams with high transverse k components as it is dominated along transverse direction rather than propagation direction during the coupling with the nanostructures. The redistribution of energy often generates a tight hotspot without compressing the energy. Even standard optical structures like axicon also able to generate near field compact focal spot **[19].**

Even though different structured beams can create subdiffraction focal spots, they still cannot reach resolutions below 100 nm. Recently, vector beams have drawn significant interest because of the beam polarization of propagating beams in the spatial direction **[20, 21]**. Spatially dependent polarization in the cross section of the beam profile is crucial for determining the resolution in the subwavelength domain, as it involves orbital angular momentum (OAM). Some studies have been based on vortex beams but have not explored high-index lenses to improve the resolution. A Laguerre Gaussian mode consists of a helical wavefront associated with a quantized azimuthal phase change of the electric field which is equivalent to a doughnut shape beam. This circular phase structure is inherently related to the intensity distribution, which provides an annular shape during beam propagation. The OAM beam we applied here is only along one component i.e. linearly polarized, which makes our Ez component very weak, difficult to focus tightly to create diffraction confinement, still for computational comfort, we assigned a simpler version of this twisted beam. We tried to optimize this issue with the use of high index SIL, where high index provides shorter effective wavelength and SIL shape includes high NA beam and together it makes the confinement stronger **[22, 23]**. For vortex beam consideration of our work in this part, we achieved transverse intensity pattern with two bright spots at the SIL-air interface via our linearly polarized OAM beam incidence.

We presented a systematic study in combination with a high-index material based optical structure to achieve diffraction limited near-field spot under three different structured beam illuminations (Gaussian beam, dark field and Laguerre Gaussian (LG) beam). We demonstrate a design concept supported by simulation results based on an LG beam with a diamond e-SIL structure, which provides superresolution below 100 nm. We also utilize an alternative structured beam, i.e., the Hermite Gauss (HG) beam, to improve the near field confinement in the subwavelength region, as this also stimulates spatial components of the beam to achieve highly resolved images. Our numerical analysis confirms that our HG beam with a specified mode is able to generate a subwavelength focal spot in association with a nano SIL with a higher refractive index. The use of different structured beams with nanostructures provides excellent performance in superresolution imaging at the interface between the nanostructure and air, which has a wide range of applications ranging from quantum sensing [**24**] to biological imaging [**25**].

We also made a study to modify the geometry by introducing a cone, cylinder and pyramidal shaped structures to enhance the overall system performance by obtaining an FWHM value of 27 nm. By varying the dimensions of the geometry and the different beam waists, we tested the robustness of the proposed model. These nanostructures can be utilized as nanorange lenses with modified focal lengths and wider numerical apertures for near-field optical imaging.

Even though, there are many applications which deals with far field patterns at the focal plane of the beam, we extensively use the near electric field distribution at the e-SIL and air interface in order to utilize these sub-diffraction spot for imaging applications. Our designs are conceptualized and computed from experimental fabrication point of view which made our proposed designs more realistic and feasible towards practical implementation. Experimentally, e-beam lithography and high power ICP-RIE techniques can be utilized to fabricate such diamond based nano SIL or nanolens structures and these structures can be incorporated in to an AFM tip to scan over any target sample for imaging applications. In addition to established sub-diffraction techniques such as PALM [**26**], STED [**27**], hyperlenses [**8**] and super oscillatory lenses [**28, 29**], present structured beam assisted high index nanostructures provide an alternative route to achieve strong near field confinement beyond the conventional diffraction limit.

## Results and discussion

### Preliminary considerations

Superresolution can be achieved either by considering a high-index substrate with evanescent wave excitation or placing subwavelength nanostructures in close proximity to the scanning object. Here, we consider a high-index solid immersion lens to construct a sub-diffraction focal spot at the plane surface of the lens. Coherent monochromatic incident light source illumination with different beam profiles is realized through our SIL structure to achieve superresolution in the subwavelength domain. Figure 1(a) depicts a schematic design of our simulation scheme. Even though two different types of geometries (e-SIL and cone) are considered, major attention is given to SIL-based nanolens structures, which are described with structural modifications for determining resolutions below 100 nm. Indeed, this high-index material (diamond) results in low loss and a more confined beam profile inside the optical structure by improving the overall system performance in terms of resolution. A high-index e-SIL improves the beam focus inside the lens structure because of the high contrast in the refractive index value at the air–SIL interface. To enumerate a slight shift in the focal spot due to the change in the refractive index after placing the lens structure, we consider the optimized position of the nanolens in such a way that it coincides with the focal spot of the incident Gaussian beam. Furthermore, the numerical aperture also simultaneously increased because of the curved surface of the SIL outer interface.

Special attention is given to the bottom surface of the e-SIL and the cone structure, as the field distribution is calculated at that plane to measure the resolution strength after the incident beam converges at the focal plane inside the high-index SIL structure. This plane is kept as a reference while scanning any sample via our nanoscale structures; hence, it is commonly known as the plane of interest (POI). The near-field distribution is the basis of subwavelength imaging when we deal with structured beam illumination, and indeed, the propagation plane of each incident beam plays a significant role in orthogonally intersecting the POI at the plane surface of the optical structure.

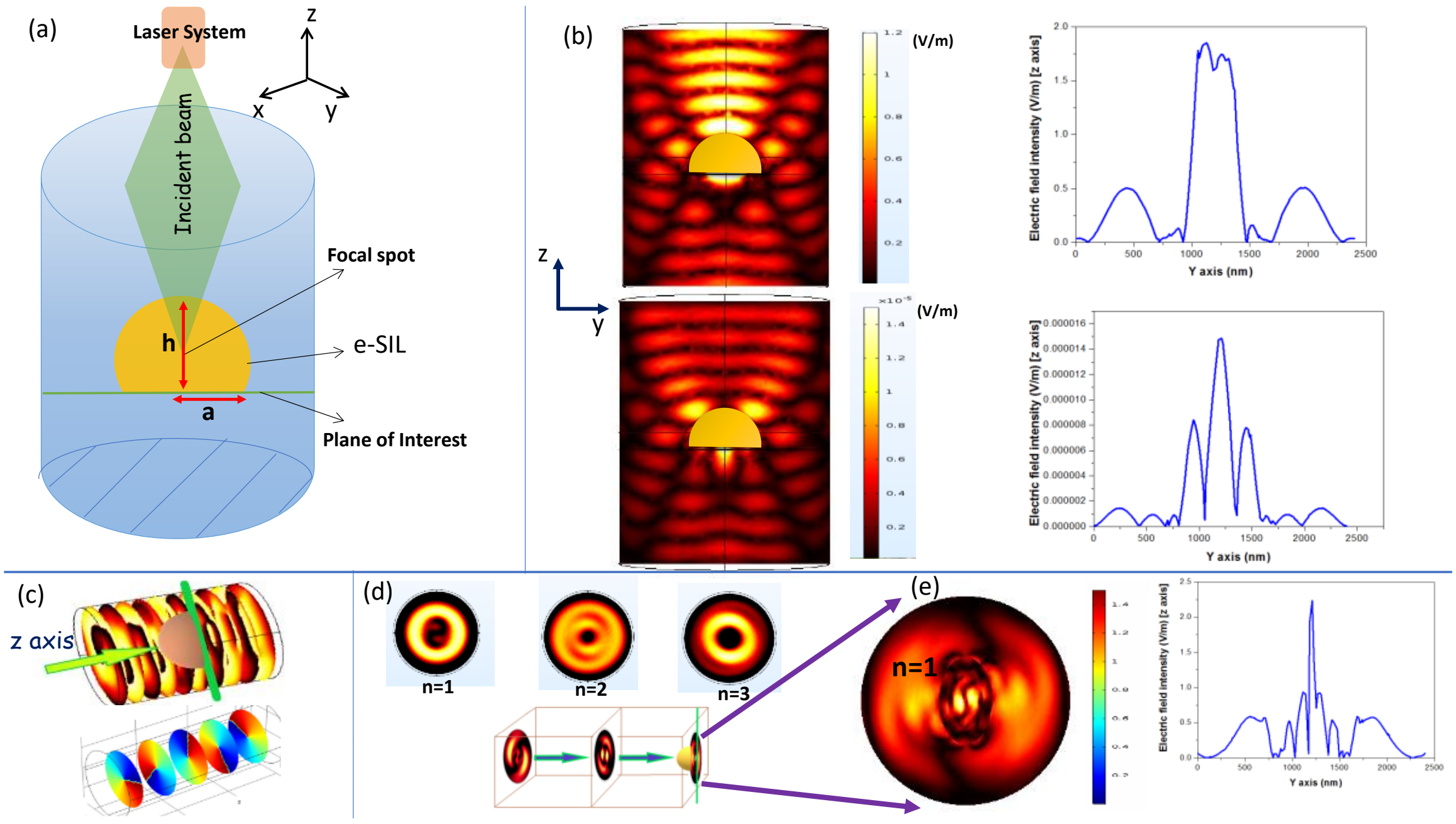

Fig. 1: (a) Schematic diagram of the optical system for the superresolution scheme. (b) Electric field distribution while the plane Gaussian beam and dark field are incident on the diamond-based ellipsoid optical nanostructure. (c) Laguerre Gaussian beam is launched toward the z-axis from a circular aperture. (d) Different modes of the LG beam are shown, we work with the optimized mode n=1, i.e., the basic mode of the LG beam. (e) Cross-sectional view of the near-field distribution at the plane of interest after SIL is placed. The near-field distribution is achieved with an FWHM of 60 nm.

To determine the FWHM value of the intensity pattern, we focus on the near electric field distribution, which is intrinsically linked with the resolution, i.e., the smallest value of the FWHM yields the highest resolution in the lateral direction. In general, we calculate a normalized distribution of the near electric field pattern ($|E^2|$) at x=0 on the y-axis plane from the POI to calculate the full width at half maximum (FWHM) value. We address a numerical calculation to mimic the experimental environment by considering a cylindrical simulation area. Figure 1(a) evidently presents the beam propagation pathway and coupling with SIL nanostructure scheme. Different incident beams with a circular aperture of 700 nm are focused on the nanoscale lens structure (made of diamond material) via the cylindrical geometry shown in this figure.

We consider an ellipsoid SIL here instead of spherical SIL, as we have three-dimensional ellipsoid structure, we have three degrees of freedom to control the optimization. For this study, as we are mainly concentrated along YZ axis to find the spatial confinement, the radius and the vertical height is being tracked to achieve a narrower FWHM. After reaching a reasonable limit of resolution (in terms of FWHM), we vary the semi major axis (a) along Y axis by keeping the height (h) constant where we obtain a better FWHM value at a 'a' value of 500 nm when the optimized height (h) is 420 nm (shown in figure 1) under the excitation wavelength ($\lambda = 532$ nm).

From our study, it is clear that the position of the incident aperture and the focal plane of the beam cause a consequential variation in the electric field distribution. Hence, the system needs stable optimization, which produces high resolution in terms of the intensity distribution even with a slight variation in the simulation parameters.

**Gaussian beam illumination**

Initially, the analysis begins by focusing a user-defined Gaussian beam [Figure 1(b)] inside the ellipsoid diamond solid immersion lens (e-SIL) optical structure. We define the wave function for launching a plane Gaussian beam as expressed below.

$$\psi(x,y,z,t) \sim e^{\frac{-(x^2+y^2)}{\omega^2}} A_0 e^{i(kz-\omega t)} e^{\frac{ik(x^2+y^2)}{2R(z)}}$$

where $\psi$ is the complete wave function for the Gaussian wave distribution, which consists of a Gaussian beam profile, unidirectional beam propagation and wave-front curvature. $A_0$ is the amplitude of the incident beam profile, and R(z) is the radius of curvature of the beam profile. The Rayleigh range is considered as

$$Z_0 = \frac{\pi {\omega_0}^2}{\lambda}, \omega_0 = beam\ waist$$

and the curvature R(z) is defined as $z(1+\frac{{Z_0}^2}{z^2})$. The ideal focal point is kept at z=0 in the simulation coordinate system. We have ignored the negligible impact of the paraxial approximation and the shift in the focal point even after the insertion of nanostructure. Figure 1 shows the electric field intensity distribution at the POI when three different beams are incident on the optimized model. The highest resolution is achieved with the application of the Laguerre Gaussian (LG) beam, which is discussed in a later section. Figure 1(b) shows the normalized near-field distribution with gradual optimization in the side lobes when a user-defined Gaussian beam profile is launched into the focal plane; correspondingly, an FWHM value of 360 nm is achieved at the POI. The resolution strength is failed to reach a value beyond the classical diffraction limit, and the near-field distribution is accompanied by small side lobes. The contrast between the side lobes and main peak is quite large at the POI. This approach results in a wider window while scanning any sample or substrate at the POI of our proposed optical nanostructure during resolution estimation.

**Dark-field illumination**

Dark field illumination with distinctive high-contrast properties is launched via the combination of two Gaussian beams of the same size with a π phase difference. We assume a straightforward approach by considering two beam profiles with the same size and amplitude, with the exception of their phases, to make our simulation easier. By reshaping the above-mentioned Gaussian beam profile with the same focal point, we express the dark field illumination as

$$\begin{aligned}\psi(x,y,z,t) &= A_0 \frac{\omega_0}{\omega(z)} e^{\frac{-(x^2+y^2)}{\omega(z)^2}} e^{i(kz-\omega t)} e^{\frac{ik(x^2+y^2)}{2R(z)}} e^{-i\varphi(z)} \\ &+ A_0 \frac{\omega_0}{\omega(z)} e^{\frac{-(x^2+y^2)}{\omega(z)^2}} e^{i(kz-\omega t+\pi)} e^{\frac{ik(x^2+y^2)}{2R(z)}} e^{-i\varphi(z)}\end{aligned}$$

Figure 1(b) presents a significant drop in the FWHM value (175 nm) of the near-field intensity pattern, making it quite better in terms of resolution performance accompanied by dominant side lobes. The energy density of the main peak is greater than that of the side lobes, which are treated as the dominant peak for resolution estimation. We obtained a subdiffraction-limited FWHM value that is reduced by almost ~200 nm less than our Gaussian illumination in this attempt. Even though the resolution limit crosses below the diffraction limit near the electric field spot generated at the center, side lobes are still present in the intensity distribution.

**LG beam illumination**

Considering our prime requirement, we introduce a Laguerre Gaussian (LG) beam onto our high-index diamond-based e-SIL structure as a new approach for the application of twisted beams. Figure 1(c) depicts the phase distribution and the spiral electric field distribution of LG beam. Even though a dark field can cross the diffraction limit, a twisted beam profile (LG beam) is still the prevailing technique to achieve better accuracy in terms of resolution, which can reach below 100 nm and is not the case with dark field illumination. Different techniques can be adopted to form LG beams, such as spiral phase plates and multimodal crystals. The phase component exp(iφ) generates orbital angular momentum. The phase singularity in the beam axis while propagating leads to a null electric field distribution in the middle of the LG beam. This beam profile corresponds to a donut shaped or concentric ring field distribution that excites only the spatial distribution of the beam structure, which helps to obtain information in the superresolution regime (i.e., below 100 nm). This beam is focused on a curved surface of the diamond lens. The high-index lens directs the circular spatial beam profile to a converging point, which slightly moves from the original focus point. Finally, at the interface between air and lens, looking at the near-field distribution, a localized tiny focal spot in the superresolution regime, are generated.

Our key prerequisite is to implement the spiral phase front beam or vortex beam with the diamond e-SIL to converge into a subdiffraction focal spot. Generally, this beam consists of a polynomial term that defines the mode of the LG beam, i.e., $L_m^n$. Here 'm' is the radial mode number and the angular mode number is represented as 'n'. We initiate an approximated Laguerre Gaussian beam in the COMSOL simulation environment, which is represented for our particular structure via the following equation:

$$E_{0,1}(r,z) = E_0 e^{i\varphi} \frac{r}{\omega(z)} \exp\left(i\left(\psi_{0,1}(z) - \psi_{0,0}(z)\right)\right) E_{0,0}(r,z)$$

where, $\psi_{m,n}(z) = (n + 2m + 1)\arctan\left(\frac{z}{z0}\right)$. We consider a basic-order LG beam for our simulation, where the angular mode number is considered to be 1 (Figure 1(d)) for LG beams of different orders. With the presence of this beam, we attempt to inspect the improvement in resolution by our model at the same POI. The highest resolution is obtained, which is ~60 nm at its optimum position (Figure 1(e)). Even though side lobes are present but the electric field distribution is less at the POI inside those lobes, hence significant difference in the resolution is not observed. The FWHM value increased significantly under LG beam illumination compared with Gaussian and dark field illumination, respectively.

## Geometrical tolerance of the optical structure

The variation in the resolution and near-field intensity distribution at the POI with respect to the beam waist of the beam profile and the geometry of the e-SIL structure along the yz plane should be tested to verify the robustness of the complete system. The intensity distribution corresponding to a wide range of beam waist variations ranging from 700 nm to 900 nm for the Gaussian, dark field and LG beams are shown in Figure 2, while the other model parameters remain at their optimum values. We achieve a FWHM of 63 nm (λ/8.5) when the beam waist is 700 nm. The green plot shows that the variation in the FWHM value from the optimum value does not exceed ~20 nm.

Figure 2 shows a comparative study to validate that our high-index material with LG beam incidence provides a substantial improvement in resolution compared with other commonly used lens materials, such as silica (r.i.=1.46) and polymers (r.i.=1.6).

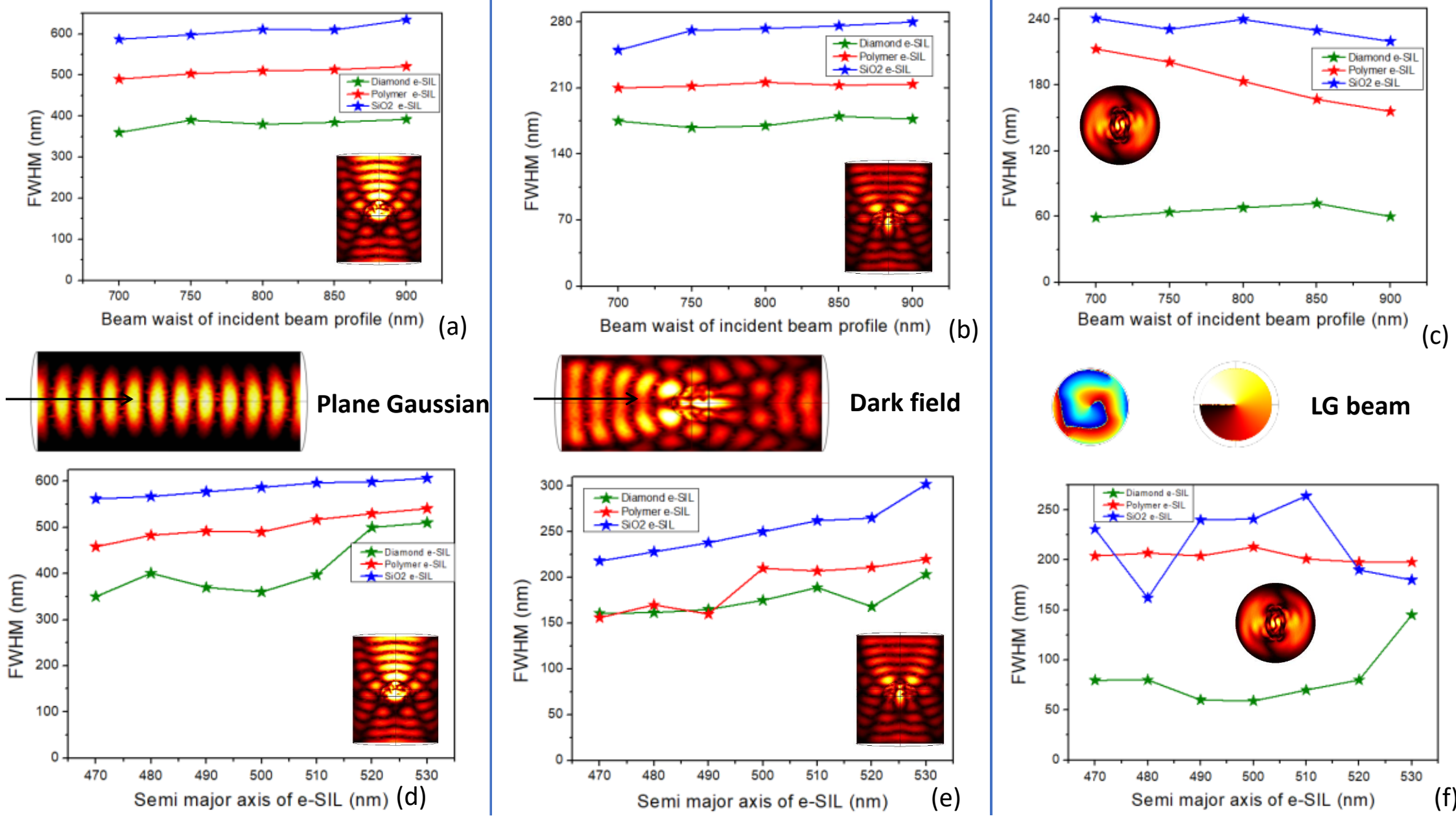

Fig. 2: Tolerance check when three different beams are launched, i.e. Gaussian, dark field and LG beams, on three different lens materials. (a) and (d) Resolution deflection is shown while the beam waist and semimajor axis are changed for the plane Gaussian beam. (b) and (e) Same analysis under dark field and (c) and (f) for the LG beam illumination.

Figure 2(a) shows a detailed comparison of three different lens materials (silica, polymer and diamond) under Gaussian beam illumination with respect to the beam waist of the incident beam. Higher beam waist indicates a minimal deviation of ±8 nm in the FWHM value, which is negligible and cannot influence the main lobe effectively. Nevertheless, the diamond-based lens produces higher resolution than the other two lens materials. Semimajor axis of the e-SIL has been varied and thus also provide poor FWHM value, as shown in Figure 2(d). The diamond lens has better resolution when the semimajor axis value is near 500 nm. The fluctuation is slightly greater for the diamond lens under Gaussian illumination than for the other two lenses.

Deflection in terms of resolution strength is further calculated under dark field incidence with respect to different beam waists for three different lens materials. The minimal deflection in the FWHM measurement (~±3 nm) is observed for the diamond and polymer lens structures (Figure 2(b)); however, for $SiO_2$, it shows ±12 nm variation due to the lower index material. Here, the converging beam pattern inside the structure significantly changes with increasing beam waist. The same tolerance is applied with respect to the geometry of the e-SIL structure (Figure 2(e)), which results in a ±20 nm deviation in the FWHM value (~resolution) for all three different lens materials. From this analysis, we can observe that the side lobe intensities are more dominant than Gaussian illumination because more energy is concentrated at the POI. The diamond lens also shows better stable performance than the other two lens materials. The fluctuation in the resolution performance over a wide range of semimajor axes is slightly greater because of the structural modification in the nanolens, which has a significant effect on beam convergence. Even though the near field focal spot exceeds the classical diffraction limit, we still cannot achieve resolution via this dark-field illumination near the superresolution regime, i.e., below 100 nm.

Furthermore, we calculate the robustness of the optimum structure under LG beam illumination with respect to same parametric sweep. This numerical study confirmed that an LG beam can yield a highly resolved image when it is incident on a diamond e-SIL at our optimized beam waist. The dominance in the side lobes proves that more energy is concentrated on those lobes at larger beam waists, as shown in Figure 2(c). This provides a negligible deflection in the resolution measurement, i.e., ±8 nm, toward a higher waist size. In the case of polymers and $SiO_2$ lenses, LG beams could not cross the barrier of the superresolution limit, which is a drawback for those materials. The poor performance in terms of the resolution of these available lens materials makes our diamond e-SIL favorable for subwavelength imaging. A shift in the focal point near the e-SIL-air interface, i.e., close to the plane of interest, can sometimes intensify the near-field spot and resolution, confirming the stability of the performance even with fluctuations.

The comparison between the FWHM when the geometry of the e-SIL is modified along the y-axis is important. We inspect the variation in the intensity profile while the semimajor axis is altered along the YZ axis for all these lenses. Figure 2(f) shows the deflection in the near-field spot under Laguerre Gaussian beam illumination, while the semimajor axes of e-SIL vary within a broad range of ±30 nm from the optimized model. In the case of other materials (silica and polymer), this geometrical tolerance results in large deviations/random fluctuations in the resolution performance compared with that of other incident beams in this case.

To determine the robustness of the resolution strength, we achieve stable variation under a plane Gaussian beam. Nevertheless, owing to the higher value (~500 nm, i.e., much closer to the incident wavelength) of the FWHM, the sub-diffraction field confinement decreases drastically. Rather, in dark-field illumination, crossing the optimized limit elucidates that the side lobe intensities are slightly enhanced by decreasing the resolving power. Figure 2(f) demonstrates truncating the lateral radius below 500 nm of e-SIL causes a reduction in resolution. In the case of LG beam application, the FWHM values decreased below 100 nm in the superresolution regime at the optimized position. These optimized geometry parameters facilitate to reach a FWHM value of 61 nm (λ/8.8), i.e., the smallest resolution measurement that one could expect in this scheme, but the sharpness of the resolution also decreases.

**Hermite Gaussian beam illumination:**

Here we introduce another structured beam i.e. Hermite- Gauss beam. This beam is composed of a solution of the paraxial Helmholtz equation, and it evolves with different modes on the basis of the polynomial functions. We have fed the general Hermite- Gaussian beam profile in our incident aperture, and the near electric field distribution is demonstrated in Figure 3(a). The beam is expressed as

$$E_{m,n} = E_0 \frac{\omega_0}{\omega(z)} e^{\frac{-(x^2+y^2)}{\omega(z)^2}} e^{i(kz)} e^{\frac{ik(x^2+y^2)}{2R(z)}} e^{-i(1+n+m)\varphi(z)} H_m\left(\frac{\sqrt{2}x}{\omega(z)}\right) H_n\left(\frac{\sqrt{2}y}{\omega(z)}\right)$$

where $H_m$ and $H_n$ are the Hermite polynomials, which have different modes depending on the m and n values. The generation of different modes, such as (00) (10), (01), (11), (12) and (02), is illustrated in Figure 3(b), which converges after falling on a high-index e-SIL structure. Owing to different field distributions, the near-field focal spot yields a sharp FWHM in the superresolution regime at the POI in the YZ plane. Different modes in the beam profile signify the spatial distribution of the beam, such as the LG beam, which also proliferates the evanescent wave to achieve more information in terms of superresolution.

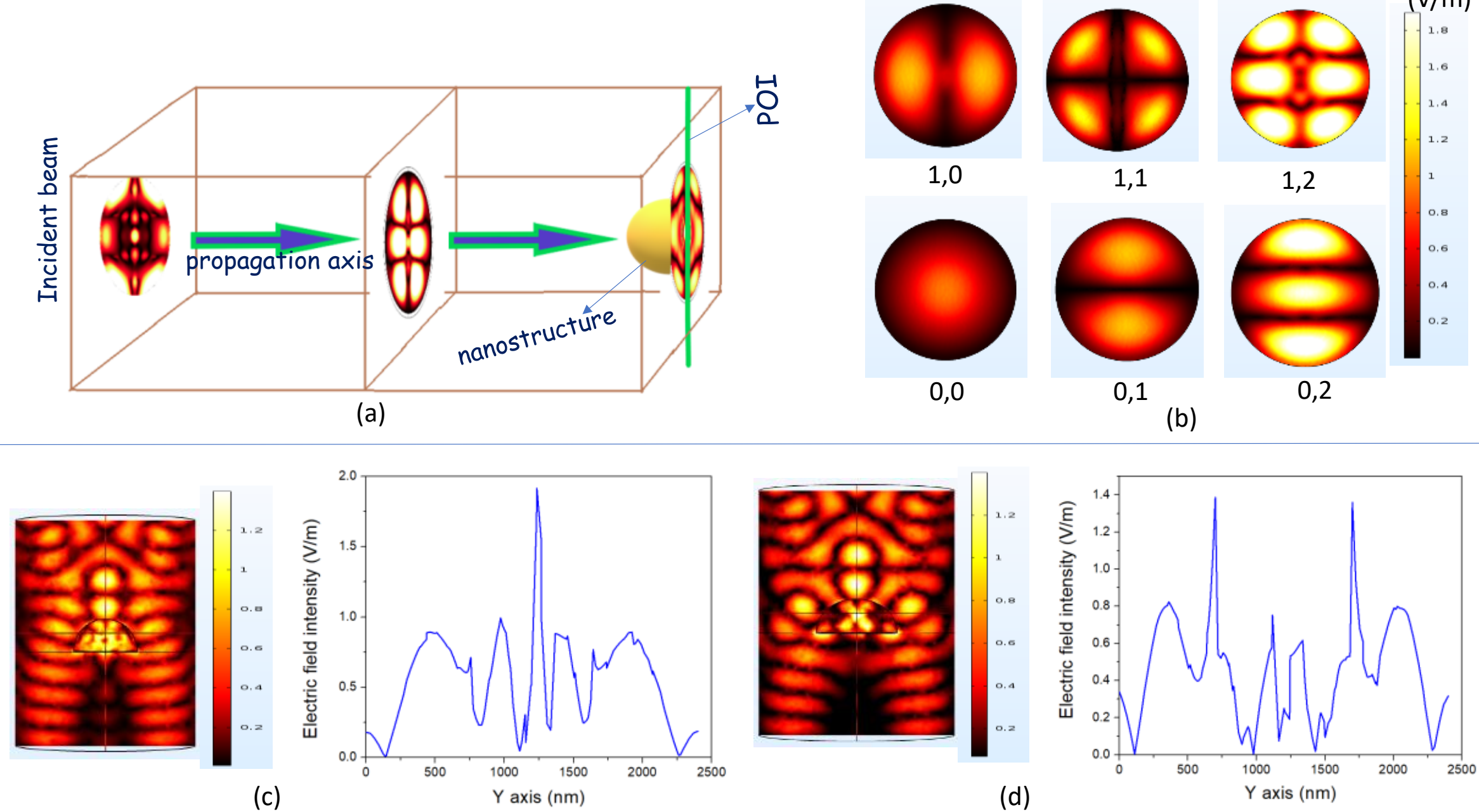

Fig 3: (a) Representation of an optical setup when a HG beam is launched on the nanorange e-SIL structure. The POI is mentioned where the near-field distribution is estimated. (b) Different higher-order modes in a HG beam are shown here for our study. (A sharp peak is achieved in the near-field distribution) $HG_{12}$ mode for our optimized structure. (d) Two distinct sharp peaks are observed for the same shape (e-SIL) with different parameters and under the $HG_{10}$ beam.

Figure 3(c) confirms that HG beam (mode 12) incidence can produce a 65 nm (~λ/8.2) superresolution image, as achieved by our high-index optimized e-SIL structure. In the case of the $HG_{10}$ mode, we obtained two distinct intensity peaks, which attained a FWHM value of 72 nm (~$\lambda$/7.4) for the diamond lens material (Figure 3(d)).

This proves that the limit of resolution in the subwavelength region can result in better performance when we use structured beam illumination on a nanoscale lens structure instead of a plane Gaussian beam. Even the results obtained in the presence of the HG beam exhibit better performance than those obtained with LG beam incidence.

We made a comparative analysis of the FWHM values of the intensity distributions of near electric field spots under plane Gaussian beam and Hermite Gaussian (HG) beam illumination after being focused by our optimized ellipsoid SIL nanostructure made from different materials (Figure 4). Even though a Gaussian beam provides a decent intensity profile in terms of the FWHM value, Hermite-Gaussian beam still employs a predominant technique to attain better accuracy in terms of resolution, which can reach even below 70 nm. To justify the choice of solid immersion lens material for the proposed model, we compared the FWHM values of three different materials, i.e., diamond, GaN and the commonly available silica (SiO2) material. Figures 4 (a) to (f) show the electric field intensity distributions at the POI when both the plane Gaussian beam and the Hermite-Gauss beam are incident on our optimized nanostructure. In that case, we observed a very large difference in the resolution when we used diamond and GaN lens materials compared with silica (Figure 4). Indeed, these high-index materials (diamond and GaN) result in low loss and a more confined beam profile, improving the overall system performance.

Finally, we obtain the highest FWHM of 63 nm ($\lambda$/8.5) under Hermite–Gauss beam illumination (incident wavelength = 532 nm) through diamond e-SIL on our optimized model (Figure 4 (f)). This explains why the Hermite–Gauss beam with a mode ($HG_{12}$) provides the best resolution while detecting any sample at the plane of interest. Even we can improve the resolution in higher-order modes in the HG beam by further optimization. We set up a HG beam with a beam waist of 750 nm to focus on our e-SIL geometry via a cylindrical aperture. When we apply the same higher-order mode ($HG_{12}$) of the user-defined Hermite–Gaussian beam, the FWHM value reaches nearly 300 nm with a commonly available silica lens, which is not under the sub-diffraction limit. Therefore, a lower index material makes the lens undesirable in the subwavelength resolution regime. Even diamond and GaN lenses also produce FWHM values near ~210 nm (~$\lambda$/2.5) in terms of resolution when this structure is exposed to a plane Gaussian beam. Therefore, Hermite-Gaussian (HG) beam illumination through the diamond e-SIL geometry is more relevant towards the implementation of superresolution techniques.

Even the same analogy as the LG beam was used here to ensure the robustness of our superresolution technique. We achieve a minimal deflection of the FWHM value from our optimum value. A higher beam waist of the incident beam helps to enhance the side lobes in the near-field distribution, which in turn decreases the resolution performance. We performed the overall study for only the high-index diamond lens structure. Figure 4(g) shows the impact of our structure on the resolution performance when the radius of the lens or semimajor axis of the lens is altered within a certain range. The shift in the resolution is almost minimal, and at higher radii, we obtain greater enhancement in the side lobes. Therefore, the resolution strength exhibits poor performance at the POI.

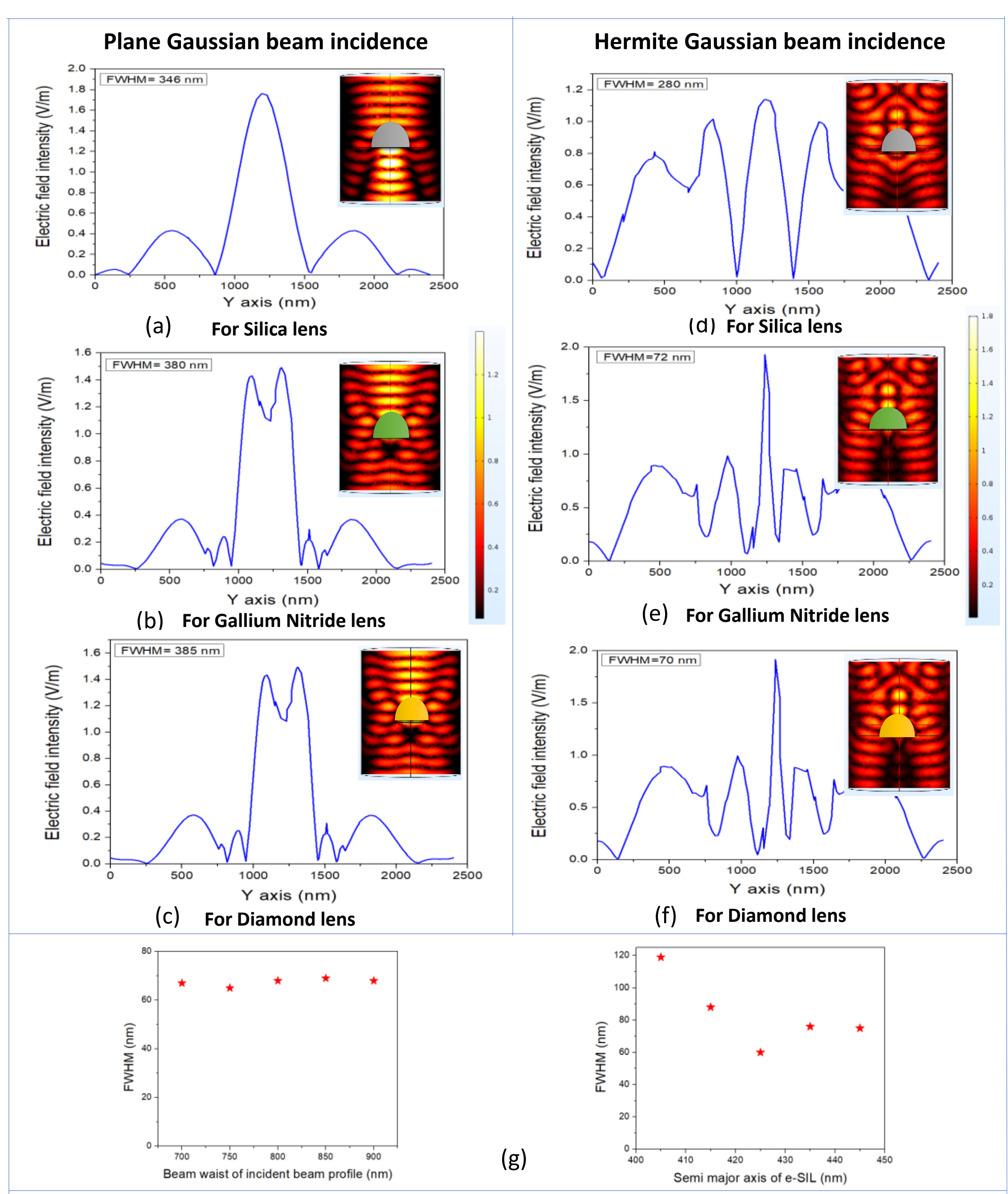

Fig. 4: (a)-(c) Plane Gaussian beam is launched on the nanorange e-SIL for three different lens materials (silica, GaN and diamond). (d)-(f) The same analysis is performed under HG beam illumination. We have shown the near electric field distribution, which is displayed in the inset. (g) This describes the impact of our structure on the resolution performance when the radius of the lens or semimajor axis of the lens is altered within a certain range.

**<u>Different structures under Laguerre- Gaussian and Hermite- Gaussian illumination:</u>**

Geometry-dependent resolution control in the subwavelength domain is an important factor in this work; therefore, we introduce several simple and unique designs that can be utilized as nanolenses to converge the beam to achieve a tiny focal spot. We introduced a high-index (diamond) material based different optical structures under the illumination of Laguerre Gaussian and Hermite Gaussian beams, which is shown in Figure 5.

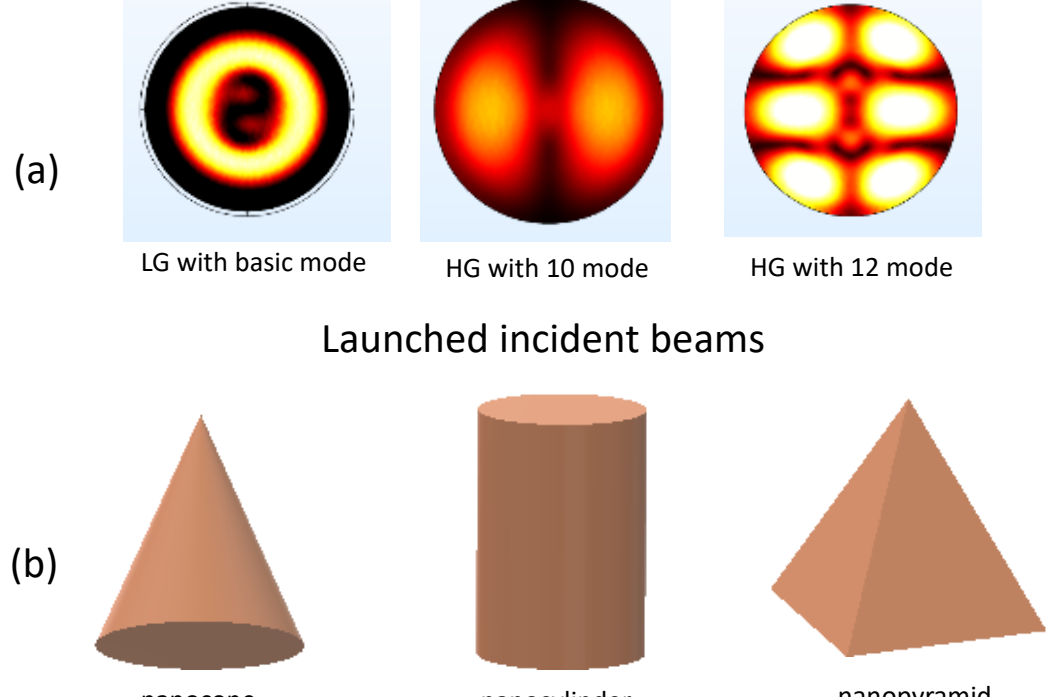

Fig. 5: (a)LG and HG beam illuminations in the presence of different optical nanostructures. We achieve our highest resolution performance in the presence of a basic LG beam and an HG beam for two modes. The beam pattern is displayed in the figure. (b) Different structures, such as nanocones, nanocylinders and nanopyramids, are used to achieve superresolution.

***<u>Geometrical analysis with Laguerre -Gaussian beam illumination:</u>***

Here we used a Laguerre- Gaussian beam from an aperture of NA=0.6 and coupled it with the nanostructures of different shapes. Later, we implemented the same analogy in this study and measured the near electric field intensity at the POI. Figure 6(a) depicts the coupling of the LG beam with the diamond-based cone structure, which generates strong near field localization at the POI. The optimized dimension of the diamond cone structure is considered as follows: the bottom radius is 430 nm, and the height is 730 nm. It is expected that a higher resolution with a lower FWHM value can be obtained with this geometry modification. We placed our optical structure at the focal plane of the incident beam. Even shifting the focal point along the z-axis significantly affects near-field spot generation. We present an optimized near electric field intensity pattern at the POI for this cone structure, which is shown in the inset (Figure 6(a)). This figure also shows that we achieved an FWHM of 57 nm (~lambda/9.5) with the LG beam incident on the cone at its optimum value. Compared with the SIL design, cone structure provides an increase in the resolution in the presence of an LG beam. In addition to improving the resolution, the cone structure also has several advantages, such as reducing the side lobe intensity, enhancing the contrast of the field distribution, and obviously shifting the focus near the POI.

We performed a geometrical sweep (the bottom radius of the cone structure was varied within ±30 nm) of the proposed cone structure for three different beams. Here, the resolution stability drastically decreases with a wide variation in the cone radius, especially for the LG beam.

Next, the tolerance limit of the resolution efficiency is studied with a variation in the cone radius (figure 6(b)) and different beam waists to make the simulation results more feasible for three different incident beams (Gaussian, dark- field and LG). Figure 6(c) shows a stable near-field intensity profile at the POI with wide beam waist variation. Moreover, for the LG beam, the dependence on the beam waist has a negligible effect on the improvement in resolution. This design offers excellent stability for the FWHM value even with a large beam waist variation.

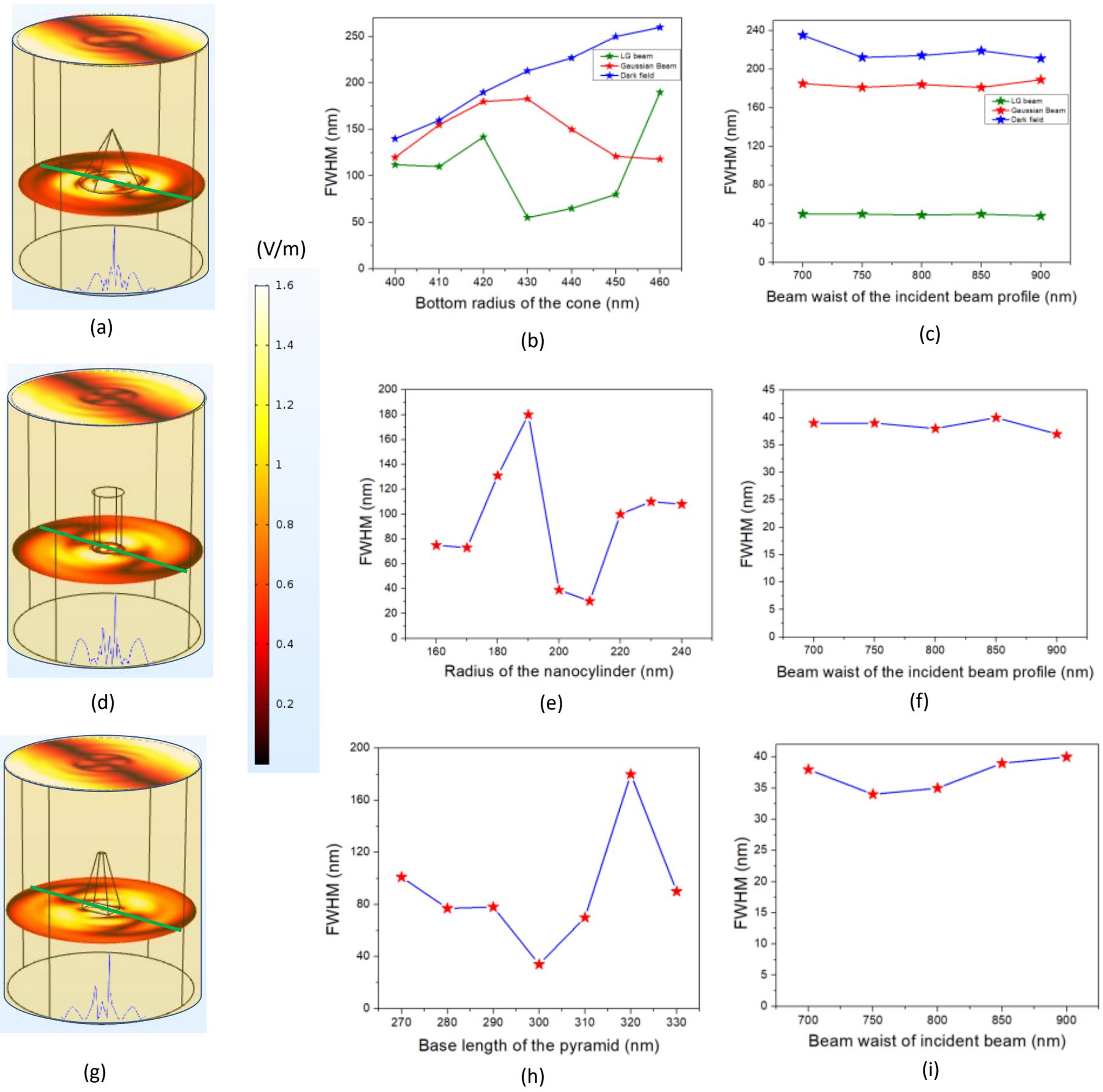

Fig. 6: (a) LG beam is launched on a nanorod cone structure. (b) (c) Geometrical tolerances of the bottom radius and resolution variation for different beam waists are displayed here for the same structure. We compared the same study in the presence of a plane Gaussian beam and dark field illumination for the same cone structure. (d)(e)(f) The near electric field distribution at POI is shown here for the nanocylinder structure with the LG beam. The optimized intensity pattern is shown in the inset. The changes in the FWHM with respect to the radius of the cylinder and the variation in the beam waist are shown. (g) (h) (i) A detailed investigation of the electric field distribution at the POI is displayed (when a LG beam is incident on a nanoscale pyramid structure). The effects of the geometry of the pyramid and the beam waist of the incident beam on the resolution measurement are studied.

In the case of nanocylinder, as shown in Figure 6(d), to improve the efficiency of FWHM measurement in the superresolution regime, we consider a basic-order LG beam to launch from the circular aperture with the same NA, which is coupled with a high-index (r.i.=2.425) material. The near-field distribution indicates that we attain a super resolution even below 50 nm, which is much better than the earlier findings. We obtained this result with an optimized radius of 200 nm and a height of 725 nm for our proposed nanocylinder. The best resolution observed for this optimized cylinder is 39 nm in terms of the FWHM value, which is equivalent to (λ/13.5). The LG beam creates a smaller side lobe in its near-field distribution, which is feasible in this case. Two important factors (i.e., radius and the beam waist) were investigated to confirm our geometrical tolerance. We varied the radius of the nanocylinder within a range of 30 nm from our optimized model. However, from Figure 6(e), we observe the variation in the FWHM values when the radius is below the optimized parameter. In the case of the beam waist, we noticed a stable variation in the FWHM measurement, which shows better performance (resolution~ λ/14) even after increasing the beam waist up to 900 nm (Figure 6(f)).

We utilized a pyramid structure in the nanoscale range, which produces a best FWHM value for the near electric field distribution at the POI of our predicted model. We obtain an FWHM value of 34 nm, which is equivalent to the (λ/15.5) resolution (Figure 6(g)). This offers a tight focal spot with lower-intensity side lobes, which improves the resolution measurement at the POI. The geometrical tolerances of various base lengths of the nanoscale pyramid structure and the incident beam waist are investigated. We have achieved significant stability over a wide range of beam waists. Figure 6(h) shows that we achieve better stability in the superresolution regime, which is much better than the other two structures under LG beam illumination. The performance stability slightly deviates when the base length of the pyramid is increased. In the next diagram (Figure 6(i)), we studied the tolerance of the beam waist of the incident LG beam. The best resolution is attained at the optimized parameter, which is the best achieved result in this structure beam illumination.

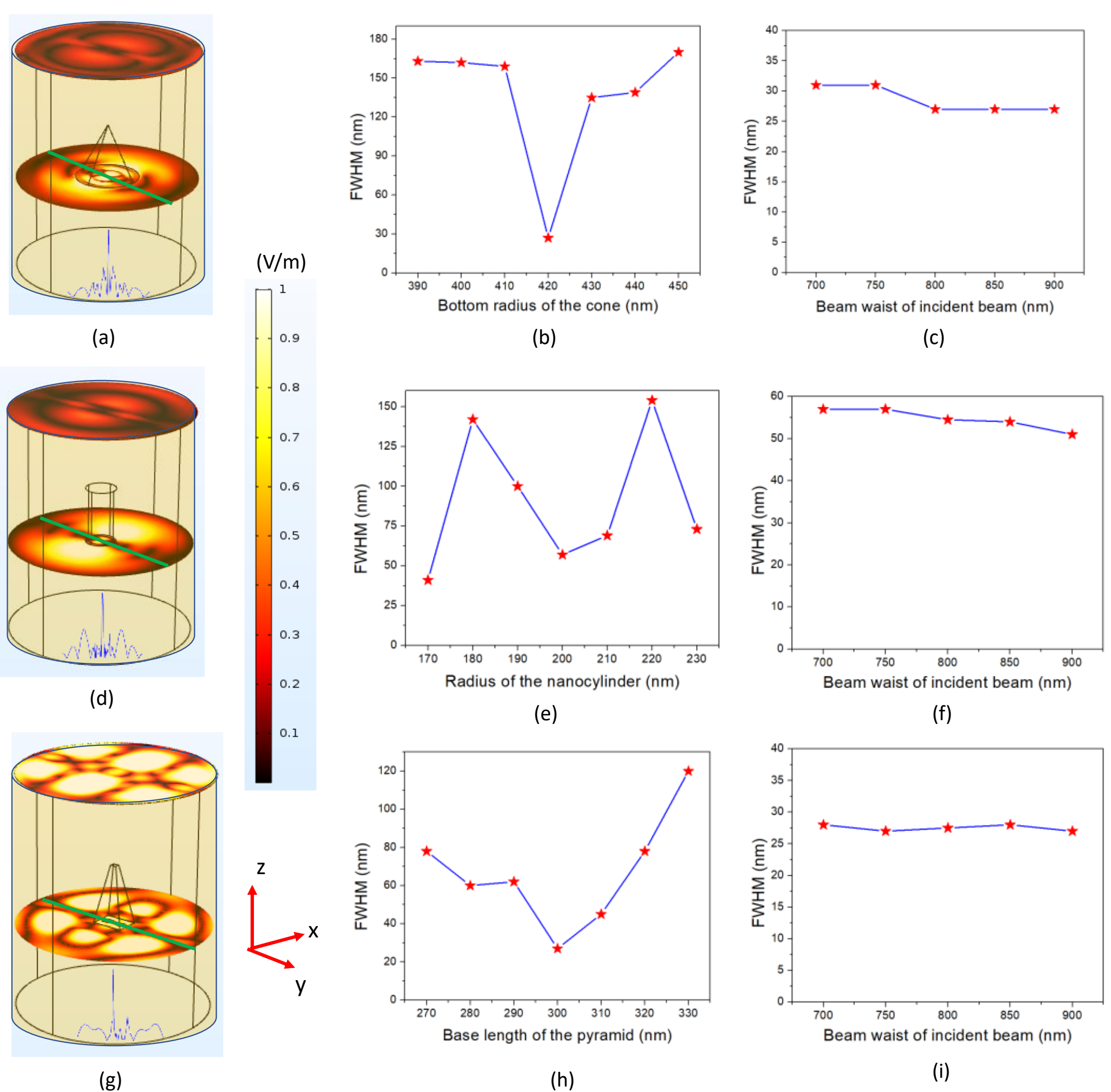

Fig. 7: (a) The HG10 beam is launched on a nanorod cone structure. (b) (c) Geometrical tolerances of the bottom radius and resolution variation for different beam waists are displayed here for the same structure. (d)The near electric field distribution at POI is shown here for the nanocylinder structure with HG10 beam illumination . The optimized intensity pattern is shown in the inset. (e)(f) The changes in the FWHM with respect to the radius of the cylinder and the variation in the beam waist are shown. (g) A detailed investigation of the electric field distribution at the POI is displayed (when a nanoscale pyramid structure is present with HG12 beam illumination). (h)(i) The effect on resolution measurement with respect to geometry of the pyramid and the beam waist of the incident beam is studied.

***Geometrical analysis with Hermite -Gaussian beam illumination:***

Here, we launched a Hermite- Gaussian beam (HG10 and HG12 modes) to all three different optical structures. The optimized results are shown in figure 7. Figures 7(a), (d), and (g) present the incident beam mode at the aperture with an NA=0.6 and the near-field distribution at the POI of three different optical structures, which can provide resolution measurements implicitly. Here, the presence of a side lobe in the near-field distribution is almost negligible for our optimized parameters. We optimized these results based on bottom radius of the cone and beam waist of incident beam (figures 7(b) and 7(c)). FWHM value shifted above the superresolution regime, which proves that our system is highly stable only at the optimized parameters. For HG10 mode incidence, diamond-based cone-shaped optical structure provides the best resolution. Coupling between the HG 10 mode incident beam and the cylindrical diamond-based optical nanostructure is shown in Figure 7(d). Here, we cannot achieve better resolution, unlike other structures, but indeed, it also has an FWHM value of 57 nm (λ/9.3). In this case, the radius of the nanocylinder exhibits more deflection when it deviates from the optimized parameter (figure 7(e)). Rather, our optical structure displays significant improvement when we move toward the larger beam waist of the incident beam (figure 7(f)). The near-field distribution shows a slightly higher side lobe with lower intensity and not much dominance, which does not greatly affect our sharp peak at the optimized parameter.

Finally, we maintained our nanoscale pyramid structure under HG12 mode beam illumination through the same NA=0.6 aperture.

We have plotted the electric field distribution where the base length of the pyramidal structure (along Y axis) is varied to check the deviation in the FWHM values. The other base length (along x axis) value is kept as constant after reaching an optimized value.

Pyramid structure can generate a tiny focal spot at the POI with a smaller side lobe field distribution. We reached a superresolution of up to 27 nm in terms of FWHM values, which is equivalent to (λ/19.5) with the optimized parameters, which is more stable when we test the geometrical tolerance of the structure. When the base length of the pyramid (along Y axis) is below its optimized value, it provides better efficiency under the superresolution regime. However, near the higher base length of the pyramid (figure 7(h)), there is a slight increase in the FWHM value, which is far from the superresolution limit. In the case of beam waist variation (figure 7(i)), it provides better stability than the other three structures. This provides evidence that improving the superresolution technique with different geometries could modify the beam profile characteristics.

**Numerical Methos and optimization process**

The simulation is studied via a Finite element method using wave optics module and frequency domain in COMSOL server with sufficiently refined meshing. We used a scattering boundary condition as an absorptive layer where there will not be any reflected (outcoming) beam from the simulation boundary which will superimpose and distort the near field localization by the nanolens structure. The maximum mesh size is ~λ/6 nm and the minimum mesh size is ~λ/100 nm which provides decent convergence of our electric field solutions. We also focus on near field localization rather than long distance wave propagation, where the numerical dispersion is almost negligible. Also, we have refined the maximum mesh size up to λ/10 to validate our mesh convergence and there is minimal shift (i.e. ±10 nm to ±15 nm) in the FWHM which confirms the subwavelength focal spot generation. We start our optimization considering a device dimension almost equivalent to the operating wavelength inside a wave optics module and continuously tracking the dimension until we achieve our desired FWHM value (~ below 100 nm). In addition, we checked different modes of the HG beam after we got a better result in the fundamental mode (plane Gaussian), where we have performed a complete set of geometrical parameter variation for each order in search of best optimized result. We start our optimization by varying any one of the parameters out of three parameters (for SIL) as we have three degrees of freedom in this 3D model builder. During the sweep over different parameters, we track which degree of freedom has a better control over the near field focal spot. It is an optimization process of three-dimensional geometrical sweep and tracking of those results to finalize the best outcome.

**CONCLUSIONS:**

In summary, we performed a numerical analysis for different geometries (e-SIL, cone, cylinder and pyramid) under different types of incident beams, i.e., Gaussian, dark field, Laguerre- Gaussian and Hermite- Gaussian beam illuminations. The near electric field intensity pattern at the plane of interest (bottom surface of the designs) is computationally investigated and analyzed. We also compared the FWHM values for different beam waists and finally obtained values of 34 nm and 27 nm under LG and HG beam illuminations for our optimized pyramidal nanostructures, which provides a superresolution over the classical diffraction limit. The proposed techniques can be utilized to cover a wide range of applications, ranging from the scanning of objects to the nanorange lithography technique, with the help of this sharp narrow intensity profile. In addition, we can even incorporate this technique with many other superresolution techniques for nanoscale imaging with extreme precision. This study could enhance the stability and practicality of the overall nanoimaging process while scanning any biological sample.